\documentclass[final,5p,times,twocolumn]{elsarticle}
\biboptions{sort&compress}
\usepackage{epsfig}

\usepackage{amssymb}
\usepackage{lipsum}
\usepackage{amsthm}
\usepackage{amsmath}
\usepackage[colorlinks=true, linkcolor=blue, citecolor=blue, urlcolor=blue]{hyperref}
\usepackage{lineno}

\usepackage{changes}

\journal{Nuclear Physics A}

\begin{document}

\begin{frontmatter}
\title{A global potential constrained by the Bohr-Sommerfeld quantization condition for $\alpha$-decay half-lives of even-even nuclei}

\author[1]{Nguyen Gia Huy}
\author[1]{Do Huy Tho}
\author[1]{Mai Doan Quang Huy}
\author[1]{Nguyen Le Anh\corref{cor1}}
\cortext[cor1]{Corresponding author}
\ead{anhnl@hcmue.edu.vn}
\affiliation[1]{organization={Department of Physics, Ho Chi Minh City University of Education},
            addressline={280 An Duong Vuong, Cho Quan Ward}, 
            city={Ho Chi Minh},
            country={Vietnam}}

\begin{abstract}
    The $\alpha$ decay provides valuable constraints on nuclear structure and plays an essential role in identifying heavy and superheavy nuclei. We study $\alpha$-decay half-lives of 178 even-even nuclei within a semi-classical WKB framework using a phenomenological Woods-Saxon $\alpha$-nucleus potential. The potential depth is determined by imposing the Bohr-Sommerfeld quantization condition (BSQC), ensuring a physically consistent description of the quasibound $\alpha$-daughter system. To facilitate large-scale calculations, a global parametrization of the BSQC-constrained potential depth is constructed. The resulting half-lives reproduce experimental data with comparable accuracy for both the direct BSQC approach and the fitted prescription, providing a first step toward a global and computationally efficient description of $\alpha$ decay.
\end{abstract}


\end{frontmatter}
\section{Introduction}
\label{sec1}

Alpha ($\alpha$) decay is one of the dominant decay modes of heavy and superheavy nuclei, representing a paradigmatic example of quantum tunneling that historically served as an early test of the Schr\"odinger equation. Owing to its clean experimental signature, $\alpha$ decay provides precise information on ground-state energies, lifetimes, and spin-parity assignments, and it remains a primary tool for the identification of new isotopes and elements~\cite{qi2019,pfutzner2012}. With more than 400 known $\alpha$ emitters and continued advances in radioactive-beam production and detection techniques, $\alpha$ decay continues to offer a sensitive probe of nuclear structure, including clustering phenomena, shell effects, effective interactions, and nuclear deformation~\cite{xu2006, deng2021, zhang2008, zhang2025, deng2020, ni2009, sun2016}.

A wide variety of theoretical approaches have been developed to describe the $\alpha$-decay process~\cite{delion2010}, ranging from fully microscopic models based on mean-field or energy-density-functional theories to macroscopic fission-like descriptions~\cite{yong2010}. In addition to these approaches, the cluster model framework offers a particularly transparent physical picture. In this framework, the parent nucleus is assumed to preform an $\alpha$ particle composed of four valence nucleons, which subsequently moves in the mean-field potential generated by the daughter nucleus~\cite{ni2009}. The decay is then described as quantum tunneling of the $\alpha$ particle through the combined nuclear and Coulomb barrier, following the seminal interpretations of Gamow~\cite{gamow1928} and of Gurney and Condon~\cite{gurney1928, gurney1929}. The penetration probability is commonly evaluated using the Wentzel-Kramers-Brillouin (WKB) semi-classical approximation~\cite{kelkar2007, chien2022,nhule2025}.

Within the WKB framework, the $\alpha$-nucleus interaction potential is a central ingredient, as it governs both the quasibound internal motion and the barrier penetrability. Microscopic $\alpha$-nucleus potentials constructed using folding models with realistic density-dependent nucleon-nucleon (NN) interactions have been shown to provide a consistent description of elastic scattering, bound-state properties, and $\alpha$ decay~\cite{chien2022,lu2024,ni2009_NPA828,xu2005}. Owing to the dominance of the real part of the interaction, $\alpha$ decay offers a particularly suitable testing ground for reducing the number of free parameters in the $\alpha$-nucleus potential. In particular, Ref.~\cite{chien2022} demonstrated that a proper implementation of the Bohr-Sommerfeld quantization condition (BSQC) within such microscopic potentials imposes stringent constraints on the $\alpha$-nucleus interaction and significantly improves the agreement between calculated and experimental $\alpha$-decay half-lives. These studies underline the predictive capability of microscopic potentials when both surface and interior regions of the interaction are reliably described.

Despite these successes, microscopic folding potentials require detailed nuclear density distributions and effective NN interactions, which can limit flexibility and increase computational cost in large-scale or systematic studies. For this reason, phenomenological $\alpha$-nucleus potentials remain widely used~\cite{qian2011,shin2016}. It is noted that both double-folding (DF) potentials and phenomenological Woods-Saxon (WS) potentials have been extensively applied as effective interactions in the analysis of nuclear reactions, scattering, and decay phenomena. When constrained by physically motivated conditions, such as reproducing the experimental $Q_\alpha$ value through the BSQC, phenomenological potentials can provide a reliable and efficient description of the quasibound $\alpha$-daughter system while retaining sensitivity to essential nuclear structure effects. However, the practical application of BSQC-constrained potentials in large-scale surveys is hindered by the need to solve semi-classical quantization conditions individually for each nucleus.

In the present work, $\alpha$ decay is investigated within the semi-classical WKB framework using a phenomenological WS $\alpha$-nucleus potential combined with the BSQC. While the BSQC provides an essential physical constraint and improves the reliability of $\alpha$-decay calculations, its direct implementation requires iterative solutions of transcendental integrals for each nucleus, rendering large-scale applications computationally demanding. To address this limitation, the WS potential depths determined from the BSQC are used to construct a fitted parametrization following the functional form proposed in Ref.~\cite{tian2024}. Unlike approaches that directly extract the potential depth from experimental half-lives, the present work determines the WS potential depth from the quasibound-state condition imposed by the BSQC, ensuring physical consistency at the wave-function level. This strategy aims to enhance numerical efficiency and consistency while reducing the complexity associated with repeated integral evaluations. The resulting $\alpha$-decay half-lives are then compared with those obtained from direct BSQC-based calculations and alternative depth-determination methods to assess the accuracy and reliability of the proposed approach. The present study is therefore intended as a first step toward a global and computationally efficient description of $\alpha$ decay based on phenomenological potentials constrained by semi-classical quantization. 

The present study is restricted to 178 even-even nuclei, for which $\alpha$ decay predominantly proceeds between the ground states of the parent and daughter nuclei. Transitions to excited states of the daughter nucleus are strongly hindered and contribute negligibly to the total decay width in systematic half-life calculations~\cite{buck1993, xu2006}. Consequently, all $\alpha$ decays considered in this work are treated as ground-state to ground-state transitions and are calculated with orbital angular momentum $L = 0$.

The remainder of this paper is organized as follows. Section~\ref{sec:method} describes the theoretical framework and computational methods, including the phenomenological potential model employed in this work, as well as the WKB approximation and the implementation of the BSQC. The calculated $\alpha$-decay results obtained using the WS potential with depths determined from both the BSQC and the proposed fitted prescription are presented and discussed in Section~\ref{sec:results}. Finally, the main conclusions are summarized in Section~\ref{sec:conclusion}.

\section{Theoretical framework and method of calculation}\label{sec:method}

\subsection{Semi-classical description for $\alpha$ decay}

The $\alpha$-decay half-life $T_{1/2}$ of a nucleus is expressed as
\begin{equation}\label{eq:T_cal}
    T_{1/2} = \frac{\ln 2}{\lambda},
\end{equation}
where $\lambda$ is the total $\alpha$-decay constant.

In the $\alpha$ cluster model, the decay constant $\lambda$ is written as the product of three factors~\cite{zhang2008, xiaojun2014},
\begin{equation}
    \lambda = S_\alpha \, \nu \, P,
\end{equation}
where $S_\alpha$ is the $\alpha$-particle preformation or spectroscopic factor, $\nu$ is the assault frequency, and $P$ is the barrier penetration probability evaluated within the WKB approximation.

In the cluster model, the $\alpha$ particle is assumed to be preformed inside the parent nucleus and to move relative to a rigid daughter core. The preformation factor $S_\alpha$ accounts for the probability of $\alpha$-nucleus formation prior to tunneling. In the present work, $S_\alpha$ is evaluated using the analytical parametrization proposed in Ref.~\cite{deng2021},
\begin{equation}\label{eq:Salpha}
    \log S_{\alpha} = a + b\left(A_d^{1/6} + A_\alpha^{1/6}\right)
    + c\frac{N}{\sqrt{Q_\alpha}}
    + d\sqrt{L(L + 1)} + h,
\end{equation}
where $A_d$ and $A_\alpha$ are the mass numbers of the daughter nucleus and the $\alpha$ particle, respectively, $N$ is the neutron number of the parent nucleus, $Q_\alpha$ is the $\alpha$-decay energy, and $L$ is the orbital angular momentum carried by the emitted $\alpha$ particle, as determined by spin-parity conservation. The coefficients $a$, $b$, $c$, $d$, and $h$ are obtained from global fits to experimental $\alpha$-decay data. Their numerical values, parameterized according to two neutron-number regions of the parent nucleus, are listed in Table~\ref{tab:S_alpha}.

\begin{table}[b]
    \centering
    \caption{Coefficients of Eq.~\eqref{eq:Salpha} for the $\alpha$-preformation factor, parameterized according to two neutron-number regions of the parent nucleus~\cite{deng2021}.}
    \label{tab:S_alpha}
    \begin{tabular}{lrrrrr}
        \hline\hline
        Region & $a$ & $b$ & $c$ & $d$ & $h$ \\
        \hline
        $N \leq 126$ & $21.232$ & $-6.210$ & $0.003$ & $-0.049$ & $0$  \\
        $N > 126$    & $37.421$ & $-10.900$ & $0.040$ & $-0.088$ & $0$ \\
        \hline\hline
    \end{tabular}
\end{table}

Within the semi-classical description, the assault frequency $\nu$ characterizes the oscillatory motion of the preformed $\alpha$ particle between the inner classical turning points and is given by~\cite{kelkar2007}
\begin{equation}\label{eq:nu}
    \nu = \frac{\hbar}{2\mu}
    \left[
    \int_{r_1}^{r_2}
    \left(
    \frac{2\mu}{\hbar^2}
    \left[Q_{\alpha} - V(r)\right]
    \right)^{-1/2}
    dr
    \right]^{-1},
\end{equation}
where $\mu$ is the reduced mass of the $\alpha$-daughter system and $V(r)$ denotes the total interaction potential.

At each encounter with the barrier, the $\alpha$ particle has a finite probability to tunnel through and escape from the nucleus. Within the WKB approximation, the barrier penetration probability $P$ is expressed as~\cite{buck1993}
\begin{equation}\label{eq:pepe}
    P = \exp\left[
    -2
    \int_{r_2}^{r_3}
    \sqrt{
    \frac{2\mu}{\hbar^2}
    \left[V(r) - Q_{\alpha}\right]
    }
    \, dr
    \right].
\end{equation}

The quantities $r_1$, $r_2$, and $r_3$ in Eqs.~\eqref{eq:nu} and~\eqref{eq:pepe} are the classical turning points defined by the condition $V(r)=Q_{\alpha}$, where $Q_{\alpha}$ is the $\alpha$-decay energy taken from Ref.~\cite{wang2021}. In practice, the inner turning point $r_1$ is very small, while the outer turning point $r_3$ can be determined approximately. Since the nuclear potential $V_N(r)$ rapidly vanishes at large distances, $r_3$ is obtained by solving the equation involving only the centrifugal and Coulomb components, $V_L(r)+V_C(r)=Q_{\alpha}$.

\subsection{Bohr-Sommerfeld quantization condition}\label{subsec:bsqc}

In the semi-classical description of $\alpha$ decay, the nuclear potential $V(r)$ must be capable of supporting a quasibound state of the $\alpha$-daughter system prior to tunneling. This requirement is fulfilled by imposing the BSQC~\cite{buck1996, xu2005},
\begin{equation}\label{eq:bsqc}
    \int_{r_1}^{r_2}
    \sqrt{\frac{2\mu}{\hbar^2}\left[Q_{\alpha} - V(r)\right]}
    \, dr
    = \left(2n + 1\right)\frac{\pi}{2},
\end{equation}
where $n = 0, 1, 2, \ldots$ are the number of nodes of the quasibound radial wave function inside the potential well. It is related to the global quantum number $G$, which characterizes the relative motion of the $\alpha$ cluster and is determined by the Wildermuth-Tang rule~\cite{wildermuth2013},
\begin{equation}\label{eq:global_qtn}
    G = 2n + L = \sum_{i=1}^{4}\left(2n_i + \ell_i\right)
    = \sum_{i=1}^{4} g_i,
\end{equation}
where $L$ is the orbital angular momentum carried by the emitted $\alpha$ particle, and $n_i$ and $\ell_i$ are the radial and orbital quantum numbers of the four nucleons forming the $\alpha$ cluster. The values of $g_i$ are chosen to enforce the Pauli exclusion principle by restricting the $\alpha$-nucleus nucleons to orbitals above the occupied shells of the daughter nucleus. In the present work, $g_i = 4$ for nucleons in the shell $50 \leq N_d, Z_d \leq 82$, $g_i = 5$ for $82 < N_d, Z_d \leq 126$, and $g_i = 6$ for nucleons outside the $N_d = 126$ shell~\cite{chien2022}. Additionally, we restrict the present analysis to the three regions $G=18$, $20$, and $22$, which mainly correspond to medium-heavy and heavy even-even nuclei where the $\alpha$-cluster picture and the BSQC-based description are expected to be most reliable~\cite{tian2024}. Under this restriction, 178 even-even nuclei are included in the systematic study, following Ref.~\cite{tian2024}. For phenomenological potentials without an explicit repulsive core, the combined use of the BSQC and the Wildermuth-Tang prescription provides an effective implementation of Pauli exclusion effects, as discussed in Ref.~\cite{chien2022}.

The BSQC thus provides a physically motivated constraint on the nuclear potential, ensuring the correct oscillatory behavior of the quasibound wave function inside the potential well and reducing ambiguities in the choice of potential depth. Its implementation is therefore essential for obtaining reliable and consistent $\alpha$-decay half-life calculations within the semi-classical framework.

\subsection{Phenomenological $\alpha$-nucleus potential}

In $\alpha$ decay, only the real part of the $\alpha$-nucleus interaction potential is relevant. Accordingly, the interaction potential employed in the present calculations consists of the nuclear, centrifugal, and Coulomb components
\begin{equation}\label{eq:pot}
    V(r) = V_N(r) + V_L(r) + V_C(r),
\end{equation}
where the nuclear part $V_N(r)$ is described by a phenomenological Woods-Saxon (WS) potential,
\begin{equation}\label{eq:wspot}
    V_N(r) = -V_0 \left[1 + \exp\left(\frac{r - R_0}{a_0}\right)\right]^{-1},
\end{equation}
where $V_0$ is the potential depth, $R_0$ is the nuclear radius of the daughter nucleus $A_d$, and $a_0$ is the diffuseness parameter. In this work, the depth $V_0$ is determined by imposing the BSQC given in Eq.~\eqref{eq:bsqc}. The centrifugal potential $V_L(r)$ accounts for the orbital angular momentum carried by the emitted $\alpha$ particle and is given by
\begin{equation}\label{eq:cfpot}
    V_L(r) = \frac{\hbar^2}{2\mu r^2}\left(L + \frac{1}{2}\right)^2,
\end{equation}
where the replacement of $L(L+1)$ by $(L+1/2)^2$ corresponds to the Langer modification~\cite{langer1937}, which ensures the correct behavior of the radial wave function near the origin and improves the validity of the WKB connection formulas~\cite{morehead1995}.

The Coulomb interaction $V_C(r)$ between the daughter nucleus and the $\alpha$ particle is modeled by assuming a uniformly charged sphere,
\begin{equation}
    V_C(r) =
    \begin{cases}
        \dfrac{Z_d e^2}{R_0}
        \left(3 - \dfrac{r^2}{R_0^2}\right), & r < R_0, \\[6pt]
        \dfrac{2Z_d e^2}{r}, & r \geq R_0,
    \end{cases}
\end{equation}
where $Z_d$ is the charge of the daughter nucleus. The radius parameter is taken as $R_0 = r_0 A_d^{1/3}$ with $r_0 = 1.27$~fm, and the diffuseness is fixed at $a = 0.64$~fm. These values correspond to the optimal parameters obtained by simultaneously varying $r_0$ and $a_0$ to minimize the rms deviation $\chi$ for the 178 even-even nuclei considered in this study. The sensitivity of the calculated results to these parameters and the optimized values are discussed in detail in Section~\ref{sec:results}.

\section{Results and discussion}\label{sec:results}

\subsection{Bohr-Sommerfeld quantization condition-constrained Woods-Saxon potential depth}\label{subsec:V0}

\begin{figure}
    \centering
    \includegraphics[width=\linewidth]{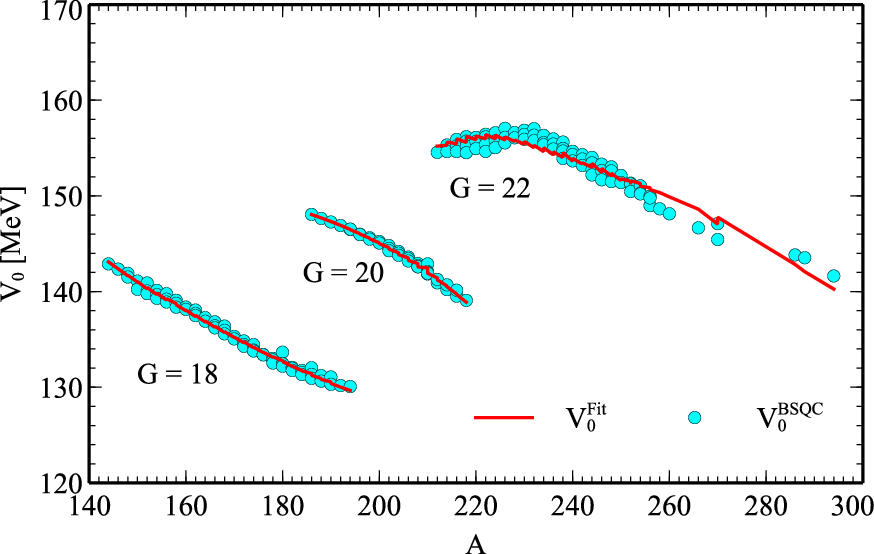}
    \caption{Depth $V_0$ of the WS potential determined from the BSQC for 178 even-even nuclei. These values are used to construct the fitted parametrization given in Eq.~\eqref{eq:V0_fit}.}
    \label{fig:V0}
\end{figure}

Figure~\ref{fig:V0} displays the depth $V_0$ of the WS potential obtained by imposing the BSQC. The results exhibit a clear separation into three distinct regions, corresponding to the three values of the global quantum number $G = 18$, $20$, and $22$ for the 178 nuclei considered. Overall, the potential depth shows a gradual decrease with increasing mass number $A$.

    \begin{table}[b]
        \centering
        \caption{Fitting coefficients of the WS potential depth $V_0$ obtained using the parametrization in Eq.~\eqref{eq:V0_fit}. All coefficients are given in MeV.}
        \label{tab:V0_para_fit}
        \begin{tabular}{crrrr}
            \hline\hline
            Region & $c_0$ & $c_1$ & $c_2$ & $c_3$
        \\  \hline
            $G = 18$ & $214.75$ & $0.02$  & $-12.04$ & $0.0004$
        \\  $G = 20$ & $47.45$ & $-0.08$ & $22.94$  & $0.0002$
        \\  $G = 22$ & $87.97$  & $-0.04$ & $15.53$  & $-0.0025$
        \\  \hline\hline
        \end{tabular}
    \end{table}

To improve computational efficiency and facilitate large-scale calculations, a fitted parametrization for the potential depth $V_0$ is introduced, following a functional form similar to that proposed in Ref.~\cite{tian2024},
    \begin{equation}\label{eq:V0_fit}
        V_0^{\mathrm{Fit}} = c_0 + \left(c_1 A^2 + c_2 A + c_3 E_{\mathrm{sh}}\dfrac{AZ}{Q_\alpha}\right)\dfrac{1}{G},
    \end{equation}
where $A$ and $Z$ are the mass and proton numbers of the parent nucleus, respectively. The quantity $E_{\mathrm{sh}}$ denotes the shell-correction energy of the parent nucleus, defined as the difference between the experimental binding energy and the macroscopic binding energy. The values of $E_{\mathrm{sh}}$ are taken from column~3 of Table~I in Ref.~\cite{tian2024}. Using the potential depths $V_0$ obtained from the BSQC, a least-squares fit is performed, and the resulting fitting coefficients are listed in Table~\ref{tab:V0_para_fit}.

The $\alpha$-decay half-lives calculated using the potential depth $V_0$ determined directly from the BSQC are compared with those obtained using the fitted parametrization in Eq.~\eqref{eq:V0_fit}. This comparison allows for an assessment of the accuracy and reliability of the fitted approach.

\subsection{Calculated $\alpha$-decay half-lives}\label{subsec:hafl-life}

\begin{figure}[t]
    \centering
    \includegraphics[width=\linewidth]{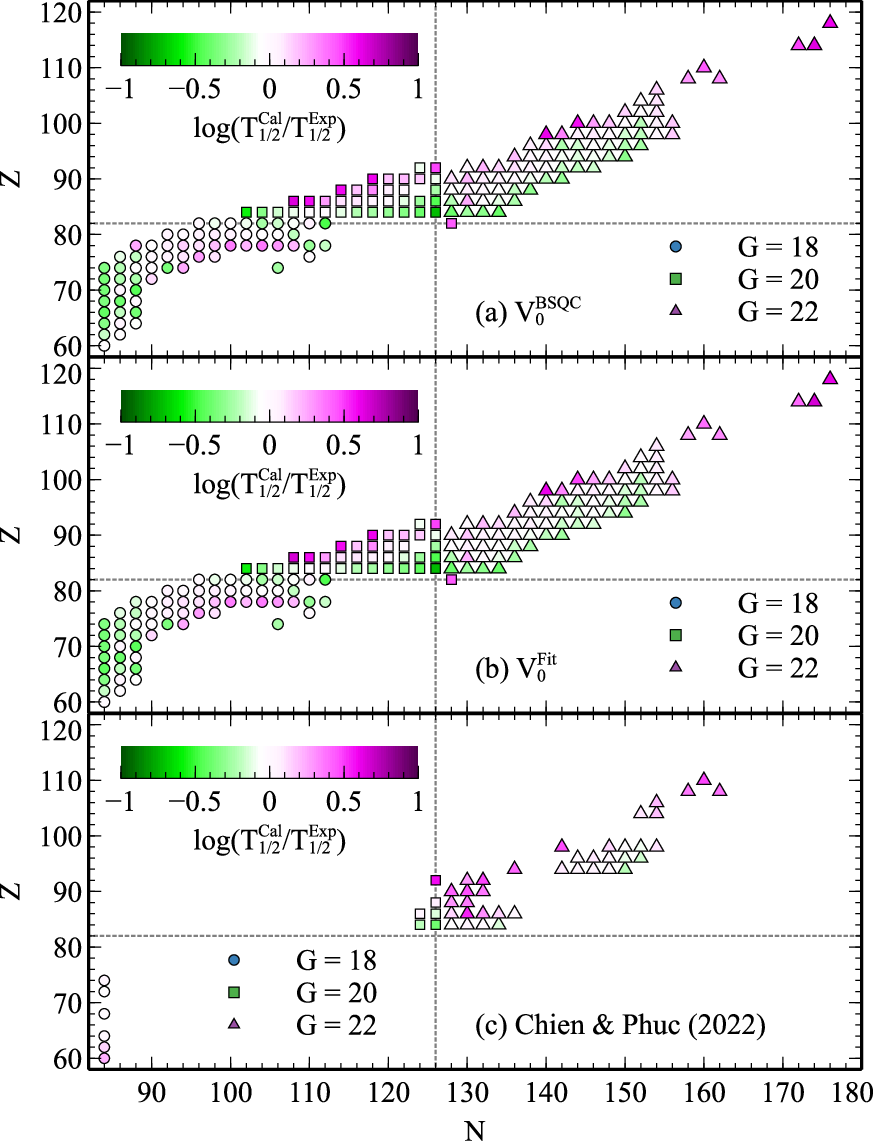}
    \caption{Nuclide-chart distribution of the decimal logarithmic deviation  $\log\left(T_{1/2}^{\mathrm{Cal}}/T_{1/2}^{\mathrm{Exp}}\right)$ between calculated and experimental $\alpha$-decay half-lives, plotted as a function of the neutron and proton numbers $(N, Z)$ of the parent nuclei. The color scale indicates the magnitude of the deviation. The results are obtained using the WS potential depth $V_0$ determined from the BSQC in Eq.~\eqref{eq:bsqc} (a) and from the fitted parametrization in Eq.~\eqref{eq:V0_fit} (b). For comparison, results calculated with the semi-microscopic DF potential based on the CDM3Y3 interaction from Ref.~\cite{chien2022} are shown in (c). The dotted lines indicate the magic numbers $N=126$ and $Z=82$. Experimental data are taken from Ref.~\cite{kondev2021}.}
    \label{fig:LogT12}
\end{figure}

To assess the accuracy of the two approaches used to determine the potential depth $V_0$, the root-mean-square (rms) deviation $\chi$ of the decimal logarithm of the calculated half-lives with respect to experimental data is evaluated as~\cite{chien2022}
    \begin{equation}
        \chi = \sqrt{\dfrac{1}{n - 1}\sum_{i=1}^n\left(\log T_{1/2, i}^{\mathrm{Cal}} - \log T_{1/2, i}^{\mathrm{Exp}}\right)^2},
    \end{equation}
where $n$ is the total number of nuclei considered.

\begin{table}[b]
    \centering
    \caption{The rms deviation $\chi$ of the calculated $\alpha$-decay half-lives $T_{1/2}$ obtained using the WS potential depth $V_0$ determined from the fitted parametrization in Eq.~\eqref{eq:V0_fit} and from the BSQC in Eq.~\eqref{eq:bsqc}. Results are grouped according to the global quantum number $G$. The second column lists the number of nuclei in each region.}
    \label{tab:chi}
    \begin{tabular}{ccrr}
        \hline\hline
        Region & $n$ & $\chi_{\mathrm{Fit}}$ & $\chi_{\mathrm{BSQC}}$
    \\  \hline
        $G = 18$ & $65$ & $0.216$ & $0.218$
    \\  $G = 20$ & $37$ & $0.341$ & $0.339$
    \\  $G = 22$ & $76$ & $0.226$ & $0.237$ \\
    \hline
    Total & $178$ & $0.250$ & $0.253$
    \\  \hline\hline
    \end{tabular}
\end{table}

Figure~\ref{fig:LogT12} compares the logarithmic deviations
$\log\left(T_{1/2}^{\mathrm{Cal}}/T_{1/2}^{\mathrm{Exp}}\right)$ obtained using the two prescriptions for determining the potential depth $V_0$. The results show that the calculated half-lives obtained with the fitted depth in Eq.~\eqref{eq:V0_fit} closely reproduce those obtained from the direct BSQC-based determination over the entire mass range $144 \leq A \leq 294$. 

Quantitatively, the rms deviation for the BSQC-based calculation is $\chi_{\mathrm{BSQC}} = 0.253$, demonstrating that the WS potential constrained by the BSQC provides a reliable description of $\alpha$-decay half-lives. When the fitted parametrization in Eq.~\eqref{eq:V0_fit} is employed, a comparable deviation of $\chi_{\mathrm{Fit}} = 0.250$ is obtained. The negligible difference between the two values indicates that the fitted prescription preserves the accuracy of the BSQC-based approach. A regional comparison of the rms deviations for nuclei grouped by the global quantum number $G$ is presented in Table~\ref{tab:chi}. In all regions, the fitted parametrization yields deviations that are comparable to, or slightly smaller than, those obtained from the direct BSQC determination.

Notably, the most pronounced improvement is observed in the $G = 22$ region, where the potential depths obtained directly from the BSQC exhibit anomalous behavior, as shown in Figure~\ref{fig:V0}. In this region, the fitted parametrization provides a smoother and more stable description of $V_0$, leading to a clear reduction of the rms deviation from $\chi_{\mathrm{BSQC}} = 0.237$ to $\chi_{\mathrm{Fit}} = 0.226$. This behavior highlights the effectiveness of the fitted prescription in mitigating local irregularities inherent in the direct BSQC-based determination. 

The agreement between calculated and experimental half-lives is more challenging for nuclei with $G=20$ than for those with $G=18$ and $G=22$. This behavior reflects enhanced nuclear-structure effects in the $G = 20$ region. The $\alpha$ decay depends not only on barrier penetrability but also on the preformation factor $S_\alpha$, which is known to exhibit significant fluctuations in transitional regions between major shell closures~\cite{varga1992PRL,varga1992NPA}. Many $G = 20$ nuclei lie in regions of shape transition and moderate deformation, where shell effects and collective correlations compete. In this regime, the nuclear structure is dominated more strongly by mean-field configurations than by well-developed $\alpha$-cluster correlations. Consequently, the $\alpha$-cluster preformation becomes more sensitive to local shell structure and deformation effects, so that a single parametrization of the WS potential depth cannot fully capture local structural variations, leading to larger rms deviations.

For comparison, calculations performed using a semi-microscopic DF potential based on the CDM3Y3 NN interaction for 50 nuclei in Ref.~\cite{chien2022} yielded an rms deviation of $\chi_{\mathrm{DF}} = 0.259$, which is close to the present value being $\chi_{\mathrm{Fit}} = 0.269$. It should be noted that in the present work, a single diffuseness parameter $a_0$ is adopted for all 178 nuclei, whereas an optimization of $a_0$ restricted to the subset of 50 nuclei could further reduce the deviation. The comparable rms deviations obtained with the WS and DF approaches demonstrate that the phenomenological WS potential provides a reasonable and competitive description of $\alpha$ decay.

From a practical perspective, the fitted potential-depth parametrization significantly reduces computational effort by avoiding repeated solutions of the BSQC integrals for each nucleus, while maintaining consistency and reliability in large-scale $\alpha$-decay calculations. Detailed numerical values for all 178 nuclei are listed in Table~\ref{tab:calc_results_full}.

\begin{table*}
    \centering
    \caption{Parameters and calculated $\alpha$-decay half-lives (in seconds). All calculations use $a_0 = 0.64$~fm and $r_0 = 1.27$~fm. The columns list: (2) the global quantum number $G$; (3-4) the shell-correction energy $E_{\mathrm{sh}}$~\cite{tian2024} and the $\alpha$-decay energy $Q_{\alpha}$~\cite{wang2021}; (5-6) the WS potential depth $V_0$ determined from the BSQC and from the fitted parametrization, respectively; (7-9) the calculated $\log T_{1/2}$ values obtained with these two prescriptions, and the corresponding experimental data~\cite{kondev2021}.} 
    \label{tab:calc_results_full}
    \begin{tabular*}{\textwidth}{@{\extracolsep{\fill}}lrrrrrrrr}
        \hline\hline
        $\alpha$ emitter & $G$ & $E_{\mathrm{sh}}$ & $Q_\alpha$ & $V_0^{\mathrm{BSQC}}$ & $V_0^{\mathrm{Fit}}$ & $\log T_{1/2}^{\mathrm{BSQC}}$ & $\log T_{1/2}^{\mathrm{Fit}}$ & $\log T_{1/2}^{\mathrm{Exp}}$ \\
        & & [MeV] & [MeV] & [MeV] & [MeV] & & & \\
        \hline
        $^{144}_{\phantom{1}60}\mathrm{Nd}$      & $18$  & $1.715$    & $1.901$    & $142.89$   & $143.10$   & $22.824$   & $22.822$   & $22.859$   \\
        $^{146}_{\phantom{1}62}\mathrm{Sm}$      & $18$  & $1.125$    & $2.529$    & $142.35$   & $142.37$   & $15.157$   & $15.156$   & $15.332$   \\
        $^{148}_{\phantom{1}62}\mathrm{Sm}$      & $18$  & $-0.049$   & $1.987$    & $141.91$   & $141.65$   & $23.248$   & $23.256$   & $23.299$   \\
        $^{148}_{\phantom{1}64}\mathrm{Gd}$      & $18$  & $0.857$    & $3.271$    & $141.52$   & $141.70$   & $9.016$    & $9.014$    & $9.350$    \\
        $^{150}_{\phantom{1}64}\mathrm{Gd}$      & $18$  & $-0.403$   & $2.807$    & $141.10$   & $140.99$   & $13.528$   & $13.529$   & $13.500$   \\
        $^{150}_{\phantom{1}66}\mathrm{Dy}$      & $18$  & $0.611$    & $4.351$    & $140.23$   & $141.05$   & $2.688$    & $2.678$    & $3.107$    \\
        $^{152}_{\phantom{1}64}\mathrm{Gd}$      & $18$  & $-1.349$   & $2.204$    & $140.90$   & $140.28$   & $21.461$   & $21.469$   & $21.533$   \\
        $^{152}_{\phantom{1}66}\mathrm{Dy}$      & $18$  & $-0.575$   & $3.727$    & $140.06$   & $140.37$   & $6.745$    & $6.741$    & $6.933$    \\
        $^{152}_{\phantom{1}68}\mathrm{Er}$      & $18$  & $0.710$    & $4.934$    & $139.79$   & $140.43$   & $0.704$    & $0.696$    & $1.057$    \\
        $^{154}_{\phantom{1}66}\mathrm{Dy}$      & $18$  & $-1.418$   & $2.945$    & $140.12$   & $139.68$   & $13.597$   & $13.603$   & $13.976$   \\
        $^{154}_{\phantom{1}68}\mathrm{Er}$      & $18$  & $-0.416$   & $4.280$    & $139.67$   & $139.76$   & $4.282$    & $4.281$    & $4.678$    \\
        $^{154}_{\phantom{1}70}\mathrm{Yb}$      & $18$  & $1.140$    & $5.474$    & $139.29$   & $139.83$   & $-0.724$   & $-0.731$   & $-0.355$   \\
        $^{156}_{\phantom{1}68}\mathrm{Er}$      & $18$  & $-1.141$   & $3.481$    & $139.77$   & $139.11$   & $9.999$    & $10.008$   & $9.989$    \\
        $^{156}_{\phantom{1}70}\mathrm{Yb}$      & $18$  & $0.041$    & $4.810$    & $139.20$   & $139.18$   & $2.432$    & $2.432$    & $2.417$    \\
        $^{156}_{\phantom{1}72}\mathrm{Hf}$      & $18$  & $2.116$    & $6.026$    & $138.91$   & $139.26$   & $-1.994$   & $-1.999$   & $-1.627$   \\
        $^{158}_{\phantom{1}70}\mathrm{Yb}$      & $18$  & $3.109$    & $4.170$    & $139.09$   & $138.75$   & $6.184$    & $6.189$    & $6.630$    \\
        $^{158}_{\phantom{1}72}\mathrm{Hf}$      & $18$  & $-0.742$   & $5.405$    & $138.77$   & $138.55$   & $0.580$    & $0.583$    & $0.810$    \\
        $^{158}_{\phantom{1}74}\mathrm{W}$       & $18$  & $0.756$    & $6.612$    & $138.37$   & $138.61$   & $-3.185$   & $-3.188$   & $-2.845$   \\
        $^{160}_{\phantom{1}72}\mathrm{Hf}$      & $18$  & $-0.125$   & $4.902$    & $138.38$   & $137.99$   & $3.050$    & $3.061$    & $3.288$    \\
        $^{160}_{\phantom{1}74}\mathrm{W}$       & $18$  & $1.877$    & $6.066$    & $138.14$   & $138.07$   & $-1.206$   & $-1.206$   & $-0.991$   \\
        $^{162}_{\phantom{1}72}\mathrm{Hf}$      & $18$  & $-0.722$   & $4.416$    & $138.07$   & $137.38$   & $5.810$    & $5.820$    & $5.690$    \\
        $^{162}_{\phantom{1}74}\mathrm{W}$       & $18$  & $0.647$    & $5.678$    & $137.70$   & $137.45$   & $0.371$    & $0.374$    & $0.420$    \\
        $^{162}_{\phantom{1}76}\mathrm{Os}$      & $18$  & $2.732$    & $6.768$    & $137.48$   & $137.52$   & $-2.789$   & $-2.799$   & $-2.680$   \\
        $^{164}_{\phantom{1}74}\mathrm{W}$       & $18$  & $-1.107$   & $5.278$    & $137.27$   & $136.81$   & $2.187$    & $2.194$    & $2.220$    \\
        $^{164}_{\phantom{1}76}\mathrm{Os}$      & $18$  & $1.777$    & $6.479$    & $136.90$   & $136.92$   & $-1.844$   & $-1.845$   & $-1.660$   \\
        $^{166}_{\phantom{1}74}\mathrm{W}$       & $18$  & $-0.550$   & $4.856$    & $136.83$   & $136.27$   & $4.375$    & $4.384$    & $4.740$    \\
        $^{166}_{\phantom{1}76}\mathrm{Os}$      & $18$  & $0.638$    & $6.143$    & $136.42$   & $136.32$   & $-0.589$   & $-0.588$   & $-0.590$   \\
        $^{166}_{\phantom{1}78}\mathrm{Pt}$      & $18$  & $2.349$    & $7.292$    & $136.20$   & $136.38$   & $-3.285$   & $-3.664$   & $-3.530$   \\
        $^{168}_{\phantom{1}74}\mathrm{W}$       & $18$  & $-0.801$   & $4.501$    & $136.39$   & $135.71$   & $6.439$    & $6.448$    & $6.199$    \\
        $^{168}_{\phantom{1}76}\mathrm{Os}$      & $18$  & $-0.044$   & $5.816$    & $135.95$   & $135.75$   & $0.741$    & $0.744$    & $0.690$    \\
        $^{168}_{\phantom{1}78}\mathrm{Pt}$      & $18$  & $1.715$    & $6.990$    & $135.58$   & $135.82$   & $-2.736$   & $-2.739$   & $-2.690$   \\
        $^{170}_{\phantom{1}76}\mathrm{Os}$      & $18$  & $-0.488$   & $5.537$    & $135.31$   & $135.19$   & $1.983$    & $1.985$    & $1.890$    \\
        $^{170}_{\phantom{1}78}\mathrm{Pt}$      & $18$  & $0.623$    & $6.707$    & $135.05$   & $135.24$   & $-1.789$   & $-1.791$   & $-1.860$   \\
        $^{172}_{\phantom{1}76}\mathrm{Os}$      & $18$  & $-0.696$   & $5.224$    & $134.84$   & $134.65$   & $3.482$    & $3.484$    & $3.210$    \\
        $^{172}_{\phantom{1}78}\mathrm{Pt}$      & $18$  & $-0.067$   & $6.463$    & $134.47$   & $134.68$   & $-0.841$   & $-0.885$   & $-0.991$   \\
        $^{172}_{\phantom{1}80}\mathrm{Hg}$      & $18$  & $1.666$    & $7.524$    & $134.28$   & $134.75$   & $-3.596$   & $-3.603$   & $-3.636$   \\
        $^{174}_{\phantom{1}76}\mathrm{Os}$      & $18$  & $-0.664$   & $4.871$    & $134.45$   & $134.13$   & $5.362$    & $5.366$    & $5.260$    \\
        $^{174}_{\phantom{1}78}\mathrm{Pt}$      & $18$  & $-0.492$   & $6.183$    & $133.89$   & $134.14$   & $0.167$    & $0.163$    & $0.061$    \\
        $^{174}_{\phantom{1}80}\mathrm{Hg}$      & $18$  & $0.642$    & $7.233$    & $133.79$   & $134.19$   & $-2.707$   & $-2.712$   & $-2.699$   \\
        $^{176}_{\phantom{1}78}\mathrm{Pt}$      & $18$  & $-0.685$   & $5.885$    & $133.42$   & $133.62$   & $1.393$    & $1.390$    & $1.199$    \\
        $^{176}_{\phantom{1}80}\mathrm{Hg}$      & $18$  & $0.051$    & $6.897$    & $133.38$   & $133.66$   & $-1.599$   & $-1.603$   & $-1.650$   \\
        $^{178}_{\phantom{1}78}\mathrm{Pt}$      & $18$  & $-0.642$   & $5.573$    & $133.01$   & $133.12$   & $2.786$    & $2.784$    & $2.430$    \\
        $^{178}_{\phantom{1}80}\mathrm{Hg}$      & $18$  & $-0.316$   & $6.577$    & $132.86$   & $133.14$   & $-0.446$   & $-0.450$   & $-0.521$   \\
        $^{178}_{\phantom{1}82}\mathrm{Pb}$      & $18$  & $0.632$    & $7.789$    & $132.53$   & $133.18$   & $-3.610$   & $-3.618$   & $-3.600$   \\
        $^{180}_{\phantom{1}74}\mathrm{W}$       & $18$  & $1.034$    & $2.515$    & $133.66$   & $132.77$   & $25.441$   & $25.454$   & $25.700$   \\
        $^{180}_{\phantom{1}78}\mathrm{Pt}$      & $18$  & $-0.424$   & $5.276$    & $132.46$   & $132.64$   & $4.245$    & $4.243$    & $4.029$    \\
        $^{180}_{\phantom{1}80}\mathrm{Hg}$      & $18$  & $-0.470$   & $6.258$    & $132.45$   & $132.64$   & $0.775$    & $0.772$    & $0.730$    \\
        $^{180}_{\phantom{1}82}\mathrm{Pb}$      & $18$  & $0.179$    & $7.419$    & $132.19$   & $132.67$   & $-2.491$   & $-2.498$   & $-2.390$   \\
        $^{182}_{\phantom{1}78}\mathrm{Pt}$      & $18$  & $-0.149$   & $4.951$    & $132.08$   & $132.18$   & $5.985$    & $5.983$    & $5.620$    \\
        $^{182}_{\phantom{1}80}\mathrm{Hg}$      & $18$  & $-0.426$   & $5.996$    & $131.98$   & $132.16$   & $1.858$    & $1.856$    & $1.890$    \\
        $^{182}_{\phantom{1}82}\mathrm{Pb}$      & $18$  & $-0.089$   & $7.066$    & $131.75$   & $132.18$   & $-1.320$   & $-1.329$   & $-1.260$   \\
        \hline
    \end{tabular*}
\end{table*}

\begin{table*}
    \centering
    \addtocounter{table}{-1}
    \caption{\textit{(Continued)}} 
    \label{tab:calc_results_full}
    \begin{tabular*}{\textwidth}{@{\extracolsep{\fill}}lrrrrrrrr}
        \hline\hline
        $\alpha$ emitter & $G$ & $E_{\mathrm{sh}}$ & $Q_\alpha$ & $V_0^{\mathrm{BSQC}}$ & $V_0^{\mathrm{Fit}}$ & $\log T_{1/2}^{\mathrm{BSQC}}$ & $\log T_{1/2}^{\mathrm{Fit}}$ & $\log T_{1/2}^{\mathrm{Exp}}$ \\
        & & [MeV] & [MeV] & [MeV] & [MeV] & & & \\
        \hline
        $^{184}_{\phantom{1}78}\mathrm{Pt}$      & $18$  & $0.285$    & $4.599$    & $131.75$   & $131.73$   & $8.082$    & $8.082$    & $7.790$    \\
        $^{184}_{\phantom{1}80}\mathrm{Hg}$      & $18$  & $-0.171$   & $5.660$    & $131.53$   & $131.70$   & $3.381$    & $3.377$    & $3.439$    \\
        $^{184}_{\phantom{1}82}\mathrm{Pb}$      & $18$  & $-0.202$   & $6.774$    & $131.32$   & $131.70$   & $-0.309$   & $-0.314$   & $-0.213$   \\
        $^{186}_{\phantom{1}76}\mathrm{Os}$      & $18$  & $1.919$    & $2.821$    & $132.04$   & $131.44$   & $22.773$   & $22.781$   & $22.800$   \\
        $^{186}_{\phantom{1}78}\mathrm{Pt}$      & $18$  & $0.825$    & $4.320$    & $131.23$   & $131.31$   & $9.943$    & $9.942$    & $9.730$    \\
        $^{186}_{\phantom{1}80}\mathrm{Hg}$      & $18$  & $0.253$    & $5.204$    & $131.34$   & $131.27$   & $5.661$    & $5.662$    & $5.710$    \\
        $^{186}_{\phantom{1}82}\mathrm{Pb}$      & $18$  & $-0.134$   & $6.471$    & $130.93$   & $131.24$   & $0.827$    & $0.823$    & $1.079$    \\
        $^{186}_{\phantom{1}84}\mathrm{Po}$      & $20$  & $-0.733$   & $8.501$    & $148.07$   & $148.05$   & $-5.049$   & $-5.049$   & $-4.470$   \\
        $^{188}_{\phantom{1}78}\mathrm{Pt}$      & $18$  & $1.455$    & $4.007$    & $130.86$   & $130.90$   & $12.244$   & $12.244$   & $12.530$   \\
        $^{188}_{\phantom{1}80}\mathrm{Hg}$      & $18$  & $0.860$    & $4.709$    & $131.21$   & $130.85$   & $8.529$    & $8.534$    & $8.720$    \\
        $^{188}_{\phantom{1}82}\mathrm{Pb}$      & $18$  & $0.201$    & $6.109$    & $130.64$   & $130.81$   & $2.304$    & $2.301$    & $2.470$    \\
        $^{188}_{\phantom{1}84}\mathrm{Po}$      & $20$  & $-0.680$   & $8.082$    & $147.64$   & $147.70$   & $-3.899$   & $-3.900$   & $-3.570$   \\
        $^{190}_{\phantom{1}78}\mathrm{Pt}$      & $18$  & $2.322$    & $3.269$    & $131.08$   & $130.56$   & $18.999$   & $19.006$   & $19.179$   \\
        $^{190}_{\phantom{1}82}\mathrm{Pb}$      & $18$  & $0.746$    & $5.698$    & $130.31$   & $130.40$   & $4.177$    & $4.175$    & $4.250$    \\
        $^{190}_{\phantom{1}84}\mathrm{Po}$      & $20$  & $-0.482$   & $7.693$    & $147.27$   & $147.32$   & $-2.770$   & $-2.771$   & $-2.611$   \\
        $^{192}_{\phantom{1}82}\mathrm{Pb}$      & $18$  & $1.539$    & $5.222$    & $130.18$   & $130.01$   & $6.609$    & $6.611$    & $6.550$    \\
        $^{192}_{\phantom{1}84}\mathrm{Po}$      & $20$  & $-0.050$   & $7.320$    & $146.90$   & $146.92$   & $-1.592$   & $-1.592$   & $-1.489$   \\
        $^{194}_{\phantom{1}82}\mathrm{Pb}$      & $18$  & $2.565$    & $4.738$    & $130.07$   & $129.67$   & $9.465$    & $9.471$    & $9.940$    \\
        $^{194}_{\phantom{1}84}\mathrm{Po}$      & $20$  & $0.559$    & $6.987$    & $146.50$   & $146.50$   & $-0.459$   & $-0.459$   & $-0.410$   \\
        $^{194}_{\phantom{1}86}\mathrm{Rn}$      & $20$  & $-0.421$   & $7.862$    & $146.53$   & $146.47$   & $-2.460$   & $-2.438$   & $-3.108$   \\
        $^{196}_{\phantom{1}84}\mathrm{Po}$      & $20$  & $1.384$    & $6.658$    & $145.99$   & $146.05$   & $0.768$    & $0.767$    & $0.777$    \\
        $^{196}_{\phantom{1}86}\mathrm{Rn}$      & $20$  & $-0.001$   & $7.617$    & $145.96$   & $146.01$   & $-1.745$   & $-1.745$   & $-2.328$   \\
        $^{198}_{\phantom{1}84}\mathrm{Po}$      & $20$  & $2.469$    & $6.310$    & $145.62$   & $145.58$   & $2.150$    & $2.151$    & $2.267$    \\
        $^{198}_{\phantom{1}86}\mathrm{Rn}$      & $20$  & $0.579$    & $7.349$    & $145.48$   & $145.52$   & $-0.880$   & $-0.881$   & $-1.159$   \\
        $^{200}_{\phantom{1}84}\mathrm{Po}$      & $20$  & $3.662$    & $5.982$    & $145.24$   & $145.10$   & $3.572$    & $3.574$    & $3.794$    \\
        $^{200}_{\phantom{1}86}\mathrm{Rn}$      & $20$  & $1.440$    & $7.043$    & $145.07$   & $145.01$   & $0.177$    & $0.178$    & $0.072$    \\
        $^{202}_{\phantom{1}84}\mathrm{Po}$      & $20$  & $5.070$    & $5.701$    & $144.81$   & $144.59$   & $4.887$    & $4.890$    & $5.143$    \\
        $^{202}_{\phantom{1}86}\mathrm{Rn}$      & $20$  & $2.489$    & $6.774$    & $144.51$   & $144.48$   & $1.186$    & $1.186$    & $1.093$    \\
        $^{202}_{\phantom{1}88}\mathrm{Ra}$      & $20$  & $0.661$    & $7.880$    & $144.27$   & $144.42$   & $-1.799$   & $-1.801$   & $-2.387$   \\
        $^{204}_{\phantom{1}84}\mathrm{Po}$      & $20$  & $6.520$    & $5.485$    & $144.19$   & $144.06$   & $5.983$    & $5.984$    & $6.276$    \\
        $^{204}_{\phantom{1}86}\mathrm{Rn}$      & $20$  & $3.632$    & $6.547$    & $144.01$   & $143.93$   & $2.067$    & $2.068$    & $2.013$    \\
        $^{204}_{\phantom{1}88}\mathrm{Ra}$      & $20$  & $1.477$    & $7.637$    & $143.79$   & $143.85$   & $-1.038$   & $-1.039$   & $-1.222$   \\
        $^{206}_{\phantom{1}84}\mathrm{Po}$      & $20$  & $8.041$    & $5.327$    & $143.60$   & $143.51$   & $6.808$    & $6.810$    & $7.146$    \\
        $^{206}_{\phantom{1}86}\mathrm{Rn}$      & $20$  & $4.875$    & $6.384$    & $143.44$   & $143.35$   & $2.726$    & $2.727$    & $2.740$    \\
        $^{206}_{\phantom{1}88}\mathrm{Ra}$      & $20$  & $2.468$    & $7.415$    & $143.20$   & $143.26$   & $-0.295$   & $-0.296$   & $-0.620$   \\
        $^{208}_{\phantom{1}84}\mathrm{Po}$      & $20$  & $9.622$    & $5.216$    & $142.92$   & $142.93$   & $7.558$    & $7.536$    & $7.961$    \\
        $^{208}_{\phantom{1}86}\mathrm{Rn}$      & $20$  & $6.144$    & $6.261$    & $142.71$   & $142.75$   & $3.251$    & $3.250$    & $3.373$    \\
        $^{208}_{\phantom{1}88}\mathrm{Ra}$      & $20$  & $3.453$    & $7.273$    & $142.59$   & $142.64$   & $0.179$    & $0.178$    & $0.107$    \\
        $^{208}_{\phantom{1}90}\mathrm{Th}$      & $20$  & $1.540$    & $8.202$    & $142.59$   & $142.57$   & $-2.022$   & $-2.022$   & $-2.620$   \\
        $^{210}_{\phantom{1}82}\mathrm{Pb}$      & $20$  & $11.360$   & $3.792$    & $142.89$   & $142.48$   & $15.006$   & $15.012$   & $14.567$   \\
        $^{210}_{\phantom{1}84}\mathrm{Po}$      & $20$  & $10.982$   & $5.408$    & $141.84$   & $142.29$   & $6.349$    & $6.344$    & $7.079$    \\
        $^{210}_{\phantom{1}86}\mathrm{Rn}$      & $20$  & $7.450$    & $6.159$    & $142.08$   & $142.11$   & $3.682$    & $3.682$    & $3.954$    \\
        $^{210}_{\phantom{1}88}\mathrm{Ra}$      & $20$  & $4.545$    & $7.151$    & $141.99$   & $141.99$   & $0.599$    & $0.599$    & $0.602$    \\
        $^{210}_{\phantom{1}90}\mathrm{Th}$      & $20$  & $2.384$    & $8.069$    & $141.90$   & $141.91$   & $-1.613$   & $-1.613$   & $-1.796$   \\
        $^{212}_{\phantom{1}84}\mathrm{Po}$      & $22$  & $8.882$    & $8.954$    & $154.55$   & $155.20$   & $-6.928$   & $-6.934$   & $-6.532$   \\
        $^{212}_{\phantom{1}86}\mathrm{Rn}$      & $20$  & $8.462$    & $6.385$    & $140.91$   & $141.43$   & $2.694$    & $2.687$    & $3.155$    \\
        $^{212}_{\phantom{1}90}\mathrm{Th}$      & $20$  & $3.204$    & $7.958$    & $141.27$   & $141.22$   & $-1.295$   & $-1.294$   & $-1.499$   \\
        $^{214}_{\phantom{1}84}\mathrm{Po}$      & $22$  & $7.048$    & $7.834$    & $155.32$   & $155.30$   & $-4.058$   & $-4.058$   & $-3.788$   \\
        $^{214}_{\phantom{1}86}\mathrm{Rn}$      & $22$  & $6.681$    & $9.208$    & $154.63$   & $155.63$   & $-6.797$   & $-6.808$   & $-6.587$   \\
        $^{214}_{\phantom{1}88}\mathrm{Ra}$      & $20$  & $6.364$    & $7.273$    & $140.21$   & $140.58$   & $0.140$    & $0.136$    & $0.387$    \\
        $^{214}_{\phantom{1}90}\mathrm{Th}$      & $20$  & $4.116$    & $7.827$    & $140.70$   & $140.51$   & $-0.896$   & $-0.894$   & $-1.060$   \\
        $^{216}_{\phantom{1}84}\mathrm{Po}$      & $22$  & $5.417$    & $6.906$    & $155.90$   & $155.44$   & $-1.164$   & $-1.159$   & $-0.842$   \\
        \hline
    \end{tabular*}
\end{table*}

\begin{table*}
    \centering
    \addtocounter{table}{-1}
    \caption{\textit{(Continued)}} 
    \label{tab:calc_results_full}
    \begin{tabular*}{\textwidth}{@{\extracolsep{\fill}}lrrrrrrrr}
        \hline\hline
        $\alpha$ emitter & $G$ & $E_{\mathrm{sh}}$ & $Q_\alpha$ & $V_0^{\mathrm{BSQC}}$ & $V_0^{\mathrm{Fit}}$ & $\log T_{1/2}^{\mathrm{BSQC}}$ & $\log T_{1/2}^{\mathrm{Fit}}$ & $\log T_{1/2}^{\mathrm{Exp}}$ \\
        & & [MeV] & [MeV] & [MeV] & [MeV] & & & \\
        \hline
        $^{216}_{\phantom{1}86}\mathrm{Rn}$      & $22$  & $5.260$    & $8.198$    & $155.23$   & $155.70$   & $-4.297$   & $-4.302$   & $-4.538$   \\
        $^{216}_{\phantom{1}88}\mathrm{Ra}$      & $22$  & $4.817$    & $9.526$    & $154.63$   & $155.97$   & $-6.808$   & $-6.800$   & $-6.764$   \\
        $^{216}_{\phantom{1}90}\mathrm{Th}$      & $20$  & $4.654$    & $8.072$    & $139.52$   & $139.74$   & $-1.673$   & $-1.676$   & $-1.580$   \\
        $^{216}_{\phantom{1}92}\mathrm{U}$       & $20$  & $2.948$    & $8.531$    & $140.14$   & $139.69$   & $-2.255$   & $-2.249$   & $-2.161$   \\
        $^{218}_{\phantom{1}84}\mathrm{Po}$      & $22$  & $4.015$    & $6.115$    & $156.18$   & $155.59$   & $1.865$    & $1.871$    & $2.270$    \\
        $^{218}_{\phantom{1}86}\mathrm{Rn}$      & $22$  & $4.010$    & $7.262$    & $155.73$   & $155.78$   & $-1.496$   & $-1.496$   & $-1.472$   \\
        $^{218}_{\phantom{1}88}\mathrm{Ra}$      & $22$  & $3.681$    & $8.540$    & $155.19$   & $156.02$   & $-4.452$   & $-4.461$   & $-4.524$   \\
        $^{218}_{\phantom{1}90}\mathrm{Th}$      & $22$  & $3.317$    & $9.849$    & $154.51$   & $156.21$   & $-6.811$   & $-6.831$   & $-6.914$   \\
        $^{218}_{\phantom{1}92}\mathrm{U}$       & $20$  & $3.342$    & $8.775$    & $139.08$   & $138.89$   & $-2.962$   & $-2.960$   & $-3.451$   \\
        $^{220}_{\phantom{1}86}\mathrm{Rn}$      & $22$  & $2.885$    & $6.405$    & $156.11$   & $155.88$   & $1.632$    & $1.636$    & $1.745$    \\
        $^{220}_{\phantom{1}88}\mathrm{Ra}$      & $22$  & $2.874$    & $7.594$    & $155.70$   & $156.01$   & $-1.735$   & $-1.739$   & $-1.742$   \\
        $^{220}_{\phantom{1}90}\mathrm{Th}$      & $22$  & $2.331$    & $8.973$    & $154.94$   & $156.27$   & $-4.779$   & $-4.818$   & $-4.991$   \\
        $^{222}_{\phantom{1}86}\mathrm{Rn}$      & $22$  & $1.933$    & $5.590$    & $156.44$   & $155.97$   & $5.284$    & $5.259$    & $5.519$    \\
        $^{222}_{\phantom{1}88}\mathrm{Ra}$      & $22$  & $2.201$    & $6.678$    & $156.17$   & $155.99$   & $1.463$    & $1.465$    & $1.526$    \\
        $^{222}_{\phantom{1}90}\mathrm{Th}$      & $22$  & $1.745$    & $8.133$    & $155.40$   & $156.23$   & $-2.571$   & $-2.581$   & $-2.650$   \\
        $^{222}_{\phantom{1}92}\mathrm{U}$       & $22$  & $1.263$    & $9.481$    & $154.65$   & $156.42$   & $-5.320$   & $-5.338$   & $-5.328$   \\
        $^{224}_{\phantom{1}88}\mathrm{Ra}$      & $22$  & $1.617$    & $5.789$    & $156.59$   & $155.95$   & $5.309$    & $5.318$    & $5.497$    \\
        $^{224}_{\phantom{1}90}\mathrm{Th}$      & $22$  & $1.430$    & $7.299$    & $155.75$   & $156.13$   & $0.063$    & $0.059$    & $0.017$    \\
        $^{224}_{\phantom{1}92}\mathrm{U}$       & $22$  & $0.826$    & $8.628$    & $155.04$   & $156.36$   & $-3.184$   & $-3.199$   & $-3.402$   \\
        $^{226}_{\phantom{1}88}\mathrm{Ra}$      & $22$  & $1.259$    & $4.871$    & $157.06$   & $155.83$   & $10.395$   & $10.461$   & $10.703$   \\
        $^{226}_{\phantom{1}90}\mathrm{Th}$      & $22$  & $1.276$    & $6.453$    & $156.12$   & $155.96$   & $3.274$    & $3.276$    & $3.265$    \\
        $^{226}_{\phantom{1}92}\mathrm{U}$       & $22$  & $0.842$    & $7.701$    & $155.54$   & $156.16$   & $-0.432$   & $-0.442$   & $-0.570$   \\
        $^{228}_{\phantom{1}90}\mathrm{Th}$      & $22$  & $1.308$    & $5.520$    & $156.61$   & $155.69$   & $7.685$    & $7.695$    & $7.781$    \\
        $^{228}_{\phantom{1}92}\mathrm{U}$       & $22$  & $1.138$    & $6.800$    & $156.07$   & $155.85$   & $2.856$    & $2.778$    & $2.748$    \\
        $^{230}_{\phantom{1}90}\mathrm{Th}$      & $22$  & $1.345$    & $4.770$    & $156.86$   & $155.40$   & $12.167$   & $12.183$   & $12.377$   \\
        $^{230}_{\phantom{1}92}\mathrm{U}$       & $22$  & $1.462$    & $5.992$    & $156.43$   & $155.47$   & $6.293$    & $6.305$    & $6.243$    \\
        $^{230}_{\phantom{1}94}\mathrm{Pu}$      & $22$  & $1.097$    & $7.178$    & $155.92$   & $155.69$   & $2.217$    & $2.220$    & $2.021$    \\
        $^{232}_{\phantom{1}90}\mathrm{Th}$      & $22$  & $1.409$    & $4.082$    & $157.02$   & $155.04$   & $17.358$   & $17.381$   & $17.645$   \\
        $^{232}_{\phantom{1}92}\mathrm{U}$       & $22$  & $1.746$    & $5.414$    & $156.34$   & $155.08$   & $9.324$    & $9.315$    & $9.336$    \\
        $^{232}_{\phantom{1}94}\mathrm{Pu}$      & $22$  & $1.545$    & $6.716$    & $155.77$   & $155.29$   & $4.013$    & $4.019$    & $4.004$    \\
        $^{234}_{\phantom{1}92}\mathrm{U}$       & $22$  & $2.010$    & $4.858$    & $156.33$   & $154.63$   & $12.714$   & $12.719$   & $12.889$   \\
        $^{234}_{\phantom{1}94}\mathrm{Pu}$      & $22$  & $1.972$    & $6.310$    & $155.57$   & $154.86$   & $5.755$    & $5.763$    & $5.723$    \\
        $^{234}_{\phantom{1}96}\mathrm{Cm}$      & $22$  & $1.776$    & $7.365$    & $155.35$   & $155.03$   & $2.363$    & $2.367$    & $2.286$    \\
        $^{236}_{\phantom{1}92}\mathrm{U}$       & $22$  & $2.042$    & $4.573$    & $155.96$   & $154.31$   & $14.675$   & $14.695$   & $14.869$   \\
        $^{236}_{\phantom{1}94}\mathrm{Pu}$      & $22$  & $2.375$    & $5.867$    & $155.42$   & $154.39$   & $7.872$    & $7.883$    & $7.955$    \\
        $^{236}_{\phantom{1}96}\mathrm{Cm}$      & $22$  & $2.216$    & $7.067$    & $154.90$   & $154.61$   & $3.485$    & $3.490$    & $3.356$    \\
        $^{238}_{\phantom{1}92}\mathrm{U}$       & $22$  & $2.003$    & $4.270$    & $155.63$   & $154.00$   & $16.990$   & $17.009$   & $17.149$   \\
        $^{238}_{\phantom{1}94}\mathrm{Pu}$      & $22$  & $2.611$    & $5.593$    & $154.95$   & $153.98$   & $9.318$    & $9.308$    & $9.442$    \\
        $^{238}_{\phantom{1}96}\mathrm{Cm}$      & $22$  & $2.746$    & $6.670$    & $154.69$   & $154.10$   & $5.082$    & $5.089$    & $5.314$    \\
        $^{238}_{\phantom{1}98}\mathrm{Cf}$      & $22$  & $1.966$    & $8.130$    & $153.92$   & $154.54$   & $0.547$    & $0.540$    & $-0.074$   \\
        $^{240}_{\phantom{1}94}\mathrm{Pu}$      & $22$  & $2.629$    & $5.256$    & $154.66$   & $153.63$   & $11.231$   & $11.245$   & $11.316$   \\
        $^{240}_{\phantom{1}96}\mathrm{Cm}$      & $22$  & $3.127$    & $6.398$    & $154.32$   & $153.63$   & $6.262$    & $6.270$    & $6.420$    \\
        $^{240}_{\phantom{1}98}\mathrm{Cf}$      & $22$  & $2.853$    & $7.711$    & $153.65$   & $153.93$   & $1.961$    & $1.947$    & $1.612$    \\
        $^{242}_{\phantom{1}94}\mathrm{Pu}$      & $22$  & $2.546$    & $4.984$    & $154.31$   & $153.32$   & $12.915$   & $12.927$   & $13.072$   \\
        $^{242}_{\phantom{1}96}\mathrm{Cm}$      & $22$  & $3.209$    & $6.216$    & $153.85$   & $153.28$   & $7.084$    & $7.091$    & $7.149$    \\
        $^{242}_{\phantom{1}98}\mathrm{Cf}$      & $22$  & $3.289$    & $7.517$    & $153.18$   & $153.46$   & $2.604$    & $2.600$    & $2.535$    \\
        $^{244}_{\phantom{1}94}\mathrm{Pu}$      & $22$  & $2.461$    & $4.666$    & $154.03$   & $152.98$   & $15.089$   & $15.102$   & $15.410$   \\
        $^{244}_{\phantom{1}96}\mathrm{Cm}$      & $22$  & $3.257$    & $5.902$    & $153.45$   & $152.89$   & $8.634$    & $8.644$    & $8.757$    \\
        $^{244}_{\phantom{1}98}\mathrm{Cf}$      & $22$  & $3.541$    & $7.329$    & $152.72$   & $153.04$   & $3.273$    & $3.269$    & $3.193$    \\
        $^{244}_{100}\mathrm{Fm}$                & $22$  & $2.966$    & $8.550$    & $152.19$   & $153.40$   & $0.073$    & $-0.017$   & $-0.506$   \\
        $^{246}_{\phantom{1}96}\mathrm{Cm}$      & $22$  & $3.273$    & $5.475$    & $153.31$   & $152.45$   & $10.946$   & $10.957$   & $11.173$   \\
        $^{246}_{\phantom{1}98}\mathrm{Cf}$      & $22$  & $3.801$    & $6.862$    & $152.64$   & $152.54$   & $5.108$    & $5.109$    & $5.111$    \\
        \hline
    \end{tabular*}
\end{table*}

\begin{table*}
    \centering
    \addtocounter{table}{-1}
    \caption{\textit{(Continued)}} 
    \label{tab:calc_results_full}
    \begin{tabular*}{\textwidth}{@{\extracolsep{\fill}}lrrrrrrrr}
        \hline\hline
        $\alpha$ emitter & $G$ & $E_{\mathrm{sh}}$ & $Q_\alpha$ & $V_0^{\mathrm{BSQC}}$ & $V_0^{\mathrm{Fit}}$ & $\log T_{1/2}^{\mathrm{BSQC}}$ & $\log T_{1/2}^{\mathrm{Fit}}$ & $\log T_{1/2}^{\mathrm{Exp}}$ \\
        & & [MeV] & [MeV] & [MeV] & [MeV] & & & \\
        \hline
        $^{246}_{100}\mathrm{Fm}$                & $22$  & $3.590$    & $8.379$    & $151.68$   & $152.86$   & $0.492$    & $0.478$    & $0.217$    \\
        $^{248}_{\phantom{1}96}\mathrm{Cm}$      & $22$  & $3.159$    & $5.162$    & $153.03$   & $152.09$   & $12.824$   & $12.835$   & $13.079$   \\
        $^{248}_{\phantom{1}98}\mathrm{Cf}$      & $22$  & $4.024$    & $6.361$    & $152.62$   & $152.00$   & $7.307$    & $7.315$    & $7.459$    \\
        $^{248}_{100}\mathrm{Fm}$                & $22$  & $3.923$    & $7.995$    & $151.50$   & $152.36$   & $1.716$    & $1.706$    & $1.538$    \\
        $^{250}_{\phantom{1}98}\mathrm{Cf}$      & $22$  & $3.986$    & $6.129$    & $152.13$   & $151.61$   & $8.422$    & $8.430$    & $8.616$    \\
        $^{250}_{100}\mathrm{Fm}$                & $22$  & $4.335$    & $7.557$    & $151.39$   & $151.79$   & $3.233$    & $3.228$    & $3.270$    \\
        $^{252}_{\phantom{1}98}\mathrm{Cf}$      & $22$  & $3.459$    & $6.217$    & $151.33$   & $151.52$   & $7.949$    & $7.947$    & $7.935$    \\
        $^{252}_{100}\mathrm{Fm}$                & $22$  & $4.655$    & $7.154$    & $151.26$   & $151.21$   & $4.757$    & $4.757$    & $4.961$    \\
        $^{252}_{102}\mathrm{No}$                & $22$  & $4.374$    & $8.549$    & $150.46$   & $151.59$   & $0.703$    & $0.689$    & $0.562$    \\
        $^{254}_{\phantom{1}98}\mathrm{Cf}$      & $22$  & $2.996$    & $5.927$    & $151.05$   & $151.30$   & $9.389$    & $9.386$    & $9.228$    \\
        $^{254}_{100}\mathrm{Fm}$                & $22$  & $4.147$    & $7.307$    & $150.27$   & $151.09$   & $4.131$    & $4.121$    & $4.068$    \\
        $^{254}_{102}\mathrm{No}$                & $22$  & $4.809$    & $8.226$    & $150.21$   & $151.00$   & $1.707$    & $1.698$    & $1.755$    \\
        $^{256}_{100}\mathrm{Fm}$                & $22$  & $3.618$    & $7.025$    & $149.98$   & $150.86$   & $5.231$    & $5.220$    & $5.064$    \\
        $^{256}_{102}\mathrm{No}$                & $22$  & $4.478$    & $8.582$    & $148.98$   & $150.81$   & $0.564$    & $0.538$    & $0.466$    \\
        $^{256}_{104}\mathrm{Rf}$                & $22$  & $5.029$    & $8.926$    & $149.78$   & $150.65$   & $0.299$    & $0.308$    & $0.328$    \\
        $^{258}_{104}\mathrm{Rf}$                & $22$  & $4.912$    & $9.196$    & $148.65$   & $150.35$   & $-0.497$   & $-0.517$   & $-0.593$   \\
        $^{260}_{106}\mathrm{Sg}$                & $22$  & $5.331$    & $9.901$    & $148.13$   & $149.89$   & $-1.628$   & $-1.678$   & $-1.767$   \\
        $^{266}_{108}\mathrm{Hs}$                & $22$  & $5.394$    & $10.346$   & $146.66$   & $148.61$   & $-2.116$   & $-2.144$   & $-2.403$   \\
        $^{270}_{108}\mathrm{Hs}$                & $22$  & $6.391$    & $9.070$    & $147.10$   & $147.05$   & $1.276$    & $1.277$    & $0.954$    \\
        $^{270}_{110}\mathrm{Ds}$                & $22$  & $5.365$    & $11.117$   & $145.44$   & $147.77$   & $-3.246$   & $-3.293$   & $-3.688$   \\
        $^{286}_{114}\mathrm{Fl}$                & $22$  & $6.502$    & $10.355$   & $143.83$   & $142.84$   & $-0.166$   & $-0.252$   & $-0.657$   \\
        $^{288}_{114}\mathrm{Fl}$                & $22$  & $6.731$    & $10.076$   & $143.53$   & $142.08$   & $0.454$    & $0.487$    & $-0.185$   \\
        $^{294}_{118}\mathrm{Og}$                & $22$  & $7.380$    & $11.867$   & $141.63$   & $140.26$   & $-2.550$   & $-2.533$   & $-3.155$   \\
        \hline\hline
    \end{tabular*}
\end{table*}

\subsection{Sensitivity to diffuseness and radius parameters}

In addition to the pronounced dependence on the global quantum number $G$, the WS potential depth $V_0$ exhibits sensitivity to the geometrical parameters of the nuclear potential, namely the diffuseness $a_0$ and the radius parameter $r_0$. To quantify this effect, the rms deviation $\chi$ was evaluated for diffuseness values in the range from $0.4$~fm to $0.8$~fm, while keeping $r_0$ fixed at $1.27$~fm.

\begin{figure}
    \centering
    \includegraphics[width=\linewidth]{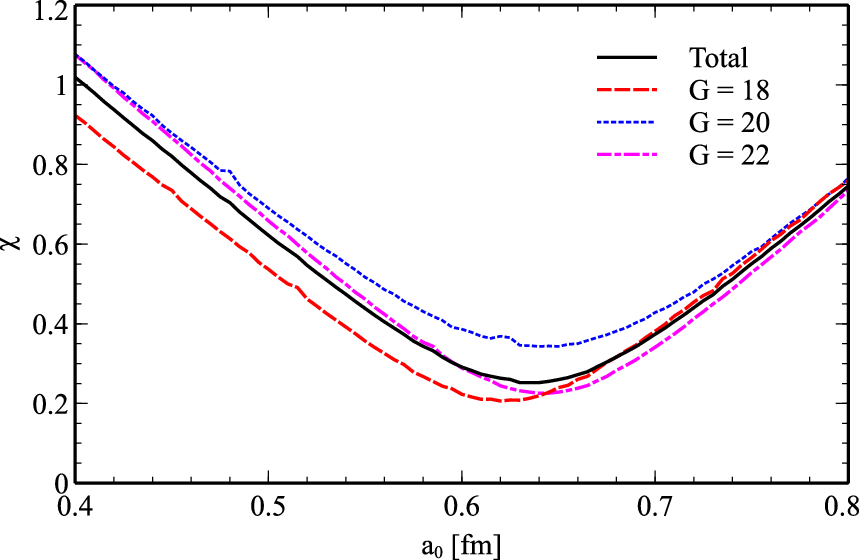}
    \caption{The rms deviation $\chi$ between the decimal logarithms of calculated and experimental $\alpha$-decay half-lives as a function of the diffuseness parameter $a_0$. The radius parameter is fixed at $r_0 = 1.27$~fm.}
    \label{fig:a0}
\end{figure}

\begin{figure}
    \centering
    \includegraphics[width=\linewidth]{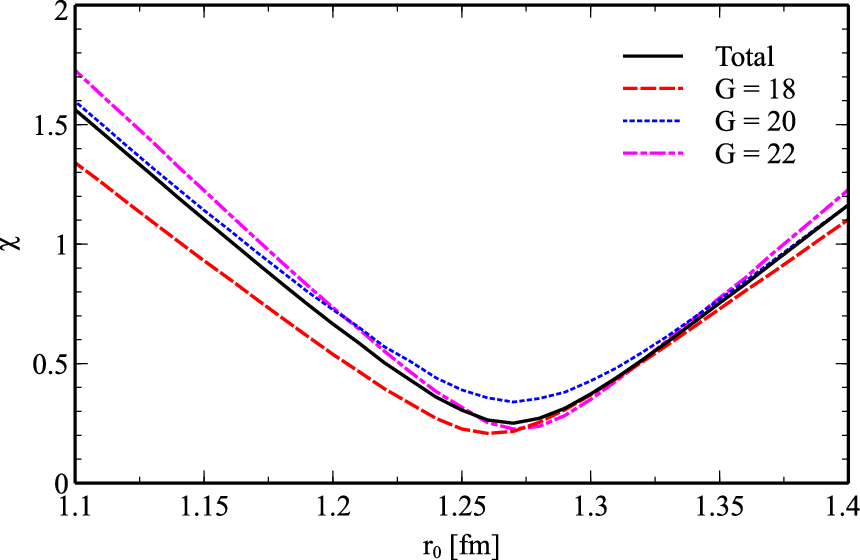}
    \caption{Same as Figure~\ref{fig:a0} but for the function of the radius parameter $r_0$. The diffuseness parameter is fixed at $a_0 = 0.64$~fm.}
    \label{fig:r0}
\end{figure}

Figure~\ref{fig:a0} shows the variation of the rms deviation $\chi$ as a function of $a_0$ for the full set of 178 nuclei. Although the dependence is not strictly monotonic, a clear minimum is observed within the investigated range, indicating the existence of an optimal diffuseness value for systematic $\alpha$-decay calculations. The optimal values of $a_0$ are found to be $0.63$~fm, $0.65$~fm, and $0.64$~fm for the regions with $G = 18$, $20$, and $22$, respectively. The average optimal value over all regions is therefore $a_0 = 0.64$~fm. When the analysis is restricted to the 50 nuclei considered, Ref.~\cite{chien2022}, the optimized diffuseness is $a_0 = 0.64$~fm, yielding an rms deviation of $\chi_{\mathrm{Fit}} = 0.251$, which is slightly smaller than the value $\chi_{\mathrm{DF}} = 0.259$ obtained using the DF approach. A regional analysis for nuclei sharing the same global quantum number $G$, also shown in Figure~\ref{fig:a0}, reveals that the $G=20$ group exhibits noticeably larger deviations than the $G=18$ and $G=22$ regions. This behavior is consistent with the stronger fluctuations of the logarithmic errors
$\log\!\left(T_{1/2}^{\mathrm{Cal}}/T_{1/2}^{\mathrm{Exp}}\right)$
around zero observed in Figure~\ref{fig:LogT12}. 

The sensitivity to the radius parameter $r_0$ is illustrated in Figure~\ref{fig:r0}, where the rms deviation $\chi$ is shown as a function of $r_0$ with the diffuseness fixed at $a_0 = 0.64$~fm. Compared to the diffuseness, the dependence on $r_0$ is smoother and less pronounced. Variations of $r_0$ primarily lead to a uniform rescaling of the potential geometry, which can be partially compensated by corresponding adjustments of the potential depth $V_0$. The optimal value $r_0 = 1.27$~fm adopted in this work is therefore determined from a global minimization of $\chi$ for the full set of 178 nuclei, providing a consistent and robust choice for systematic $\alpha$-decay calculations.

\section{Conclusion}\label{sec:conclusion}

In this work, $\alpha$-decay half-lives of 178 even-even nuclei have been calculated within the semi-classical WKB framework using a phenomenological WS potential. The novelty of the present work lies not in proposing a new decay model, but in establishing a physically constrained and computationally efficient framework that bridges direct BSQC-based calculations and global phenomenological parametrizations. When the potential depth $V_0$ is determined directly from the BSQC, an rms deviation of $\chi_{\mathrm{BSQC}} = 0.253$ is obtained, confirming the reliability of the WS potential constrained by semi-classical quantization for systematic $\alpha$-decay calculations.

Based on the BSQC-determined depths, a fitted parametrization of the WS potential depth has been constructed. The resulting half-lives yield a comparable rms deviation of $\chi_{\mathrm{Fit}} = 0.250$, demonstrating that the fitted prescription preserves the accuracy of the direct BSQC approach. From a practical perspective, this parametrization significantly reduces the computational cost by avoiding repeated semi-classical integral evaluations, while maintaining consistency and reliability in large-scale calculations.

The present study focuses on even-even nuclei, where $\alpha$ decay predominantly occurs between ground states and angular-momentum hindrance effects are minimal. In this sense, the proposed approach should be regarded as a first step toward a global and efficient description of $\alpha$ decay based on phenomenological potentials constrained by semi-classical quantization. Extensions to odd-$A$ and odd-odd nuclei, as well as the inclusion of additional structural effects such as deformation and explicit preformation dynamics~\cite{deng2016,sun2017,sun2025,denisov2024,zheng2025}, will be addressed in future work.

Overall, the fitted WS potential depth provides an efficient and accurate tool for systematic $\alpha$-decay studies, offering a practical framework for applications to heavy and superheavy nuclei.

\section*{Acknowledgements}
The authors thank Le Hoang Chien (VNU-HCMUS) and Bui Minh Loc (SDSU) for providing numerical data and for valuable discussions in this work. This research is funded by the Vietnam National Foundation for Science and Technology Development (NAFOSTED) under grant number 103.04-2025.07.

\bibliographystyle{elsarticle-num} 
\bibliography{refs}

@article{varga1992PRL,
  title = {Absolute alpha decay width of $^{212}\mathrm{Po}$ in a combined shell and cluster model},
  author = {Varga, K. and Lovas, R. G. and Liotta, R. J.},
  journal = {Phys. Rev. Lett.},
  volume = {69},
  issue = {1},
  pages = {37--40},
  numpages = {0},
  year = {1992},
  month = {Jul},
  publisher = {American Physical Society},
  doi = {10.1103/PhysRevLett.69.37}
}

@article{varga1992NPA,
  title={Cluster-configuration shell model for alpha decay},
  author={Varga, K and Lovas, R G and Liotta, R J},
  journal={Nucl. Phys. A},
  doi={10.1016/0375-9474(92)90017-E},
  volume={550},
  number={3},
  pages={421--452},
  year={1992},
  publisher={Elsevier}
}

@article{kondev2021,
doi = {10.1088/1674-1137/abddae},
year = {2021},
month = {mar},
publisher = {Chinese Physical Society and the Institute of High Energy Physics of the Chinese Academy of Sciences and the Institute of Modern Physics of the Chinese Academy of Sciences and IOP Publishing Ltd},
volume = {45},
number = {3},
pages = {030001},
author = {Kondev, F.G. and Wang, M. and Huang, W.J. and Naimi, S. and Audi, G.},
title = "{The NUBASE2020 evaluation of nuclear physics properties}",
journal = {Chin. Phys. C}
}

@article{sun2025,
  title = "{Nuclear $\ensuremath{\alpha}$ decay based on the pick-up mechanism}",
  author = {Sun, Xiaojun and Cui, Yanran and Zou, Fanglei and Xing, Kang and Tian, Junlong and Wang, Ning},
  journal = {Phys. Rev. C},
  volume = {111},
  issue = {6},
  pages = {064319},
  numpages = {9},
  year = {2025},
  month = {Jun},
  publisher = {American Physical Society},
  doi = {10.1103/ryvc-5r6x}
}

@article{deng2020,
  title = "{Analytic formula for estimating the $\ensuremath{\alpha}$-particle preformation factor}",
  author = {Deng, Jun-Gang and Zhang, Hong-Fei},
  journal = {Phys. Rev. C},
  volume = {102},
  issue = {4},
  pages = {044314},
  numpages = {17},
  year = {2020},
  month = {Oct},
  publisher = {American Physical Society},
  doi = {10.1103/PhysRevC.102.044314}
}

@article{langer1937,
  title = "{On the Connection Formulas and the Solutions of the Wave Equation}",
  author = {Langer, Rudolph E.},
  journal = {Phys. Rev.},
  volume = {51},
  issue = {8},
  pages = {669--676},
  numpages = {0},
  year = {1937},
  month = {Apr},
  publisher = {American Physical Society},
  doi = {10.1103/PhysRev.51.669}
}

@article{morehead1995,
  title="{Asymptotics of radial wave equations}",
  author={Morehead, James J},
  journal={Journal of Mathematical Physics},
  volume={36},
  number={10},
  pages={5431--5452},
  year={1995},
  publisher={American Institute of Physics},
  doi={10.1063/1.531270}
}

@article{chien2022,
    title = "{On the Bohr-Sommerfeld quantization condition and assault frequency in a semiclassical model for $\alpha$ decay}",
    journal = {Nucl. Phys. A},
    volume = {1018},
    pages = {122373},
    year = {2022},
    issn = {0375-9474},
    doi = {10.1016/j.nuclphysa.2021.122373},
    author = {Le Hoang Chien and Nguyen Tri Toan Phuc}
}

@article{kelkar2007,
  title = "{Critical view of WKB decay widths}",
  author = {Kelkar, N. G. and Casta\~neda, H. M.},
  journal = {Phys. Rev. C},
  volume = {76},
  issue = {6},
  pages = {064605},
  numpages = {8},
  year = {2007},
  month = {Dec},
  publisher = {American Physical Society},
  doi = {10.1103/PhysRevC.76.064605}
}

@article{zhang2008,
  title = "{\ensuremath{\alpha} particle preformation in heavy nuclei and penetration probability}",
  author = {Zhang, H. F. and Royer, G.},
  journal = {Phys. Rev. C},
  volume = {77},
  issue = {5},
  pages = {054318},
  numpages = {7},
  year = {2008},
  month = {May},
  publisher = {American Physical Society},
  doi = {10.1103/PhysRevC.77.054318}
}

@article{xiaojun2014,
title = "{Systematical calculation of $\alpha$ decay half-lives with a generalized liquid drop model}",
journal = {Nucl. Phys. A},
volume = {921},
pages = {85-95},
year = {2014},
issn = {0375-9474},
doi = {10.1016/j.nuclphysa.2013.11.002},
author = {Xiaojun Bao and Hongfei Zhang and Haifei Zhang and G. Royer and Junqing Li}
}

@article{buck1996,
  title = "{Cluster model of \ensuremath{\alpha} decay and $^{212}\mathrm{Po}$}",
  author = {Buck, B. and Johnston, J. C. and Merchant, A. C. and Perez, S. M.},
  journal = {Phys. Rev. C},
  volume = {53},
  issue = {6},
  pages = {2841--2848},
  numpages = {0},
  year = {1996},
  month = {Jun},
  publisher = {American Physical Society},
  doi = {10.1103/PhysRevC.53.2841}
}

@article{xu2005,
title = "{Favored $\alpha$-decays of medium mass nuclei in density-dependent cluster model}",
journal = {Nucl. Phys. A},
volume = {760},
number = {3},
pages = {303-316},
year = {2005},
issn = {0375-9474},
doi = {10.1016/j.nuclphysa.2005.06.011},
author = {Chang Xu and Zhongzhou Ren},
}

@article{wang2021,
  title="{The AME 2020 atomic mass evaluation (II). Tables, graphs and references}",
  author={Wang, Meng and Huang, Wen Jie and Kondev, Filip G and Audi, Georges and Naimi, Sarah},
  journal={Chin. Phys. C},
  volume={45},
  number={3},
  pages={030003},
  year={2021},
  doi = {10.1088/1674-1137/abddaf},
  publisher={IOP Publishing}
}

@book{wildermuth2013,
  title="{A unified theory of the nucleus}",
  author={Wildermuth, Karl and Tang, Y.C.},
  year={1977},
  publisher={Springer-Verlag},
  doi={10.1007/978-3-322-85255-7}
}

@article{xu2006,
  title = "{Global calculation of \ensuremath{\alpha}-decay half-lives with a deformed density-dependent cluster model}",
  author = {Xu, Chang and Ren, Zhongzhou},
  journal = {Phys. Rev. C},
  volume = {74},
  issue = {1},
  pages = {014304},
  numpages = {10},
  year = {2006},
  month = {Jul},
  publisher = {American Physical Society},
  doi = {10.1103/PhysRevC.74.014304},
}

@article{buck1993,
title = "{Half-Lives of Favored Alpha Decays from Nuclear Ground States}",
journal = {At. Data Nucl. Data Tables},
volume = {54},
number = {1},
pages = {53-73},
year = {1993},
issn = {0092-640X},
doi = {10.1006/adnd.1993.1009},
author = {B. Buck and A.C. Merchant and S.M. Perez},
}

@article{gamow1928,
  title="{Zur quantentheorie des atomkernes}",
  author={Gamow, George},
  journal={Z. Phys.},
  volume={51},
  number={3},
  pages={204--212},
  year={1928},
  publisher={Springer},
  doi={10.1007/BF01343196}
}

@article{gurney1928,
  title="{Wave mechanics and radioactive disintegration}",
  author={Gurney, Ronald W and Condon, Edw U},
  journal={Nature},
  volume={122},
  number={3073},
  pages={439--439},
  year={1928},
  publisher={Nature Publishing Group UK London},
  doi={10.1038/122439a0}
}

@article{gurney1929,
  title = "{Quantum Mechanics and Radioactive Disintegration}",
  author = {Gurney, R. W. and Condon, E. U.},
  journal = {Phys. Rev.},
  volume = {33},
  issue = {2},
  pages = {127--140},
  numpages = {0},
  year = {1929},
  month = {Feb},
  publisher = {American Physical Society},
  doi = {10.1103/PhysRev.33.127}
}

@article{tian2024,
  title = "{Shell correction dependence of potential depth in an $\ensuremath{\alpha}$-decay cluster model}",
  author = {Tian, Junlong and Ren, Kai and Ma, Pengfei and Li, Cheng and Wang, Ning},
  journal = {Phys. Rev. C},
  volume = {110},
  issue = {6},
  pages = {064313},
  numpages = {10},
  year = {2024},
  month = {Dec},
  publisher = {American Physical Society},
  doi = {10.1103/PhysRevC.110.064313},
}

@article{ni2009,
title = "{Microscopic calculation of $\alpha$-decay half-lives within the cluster model}",
journal = {Nucl. Phys. A},
volume = {825},
number = {3},
pages = {145-158},
year = {2009},
issn = {0375-9474},
doi = {10.1016/j.nuclphysa.2009.04.010},
author = {Dongdong Ni and Zhongzhou Ren},
}

@article{deng2021,
title = "{Correlation between $\alpha$-particle preformation factor and $\alpha$ decay energy}",
journal = {Phys. Lett. B},
volume = {816},
pages = {136247},
year = {2021},
issn = {0370-2693},
doi = {10.1016/j.physletb.2021.136247},
author = {Jun-Gang Deng and Hong-Fei Zhang},
}

@article{ni2009_NPA828,
  title="{$\alpha$-Decay calculations of medium mass nuclei within generalized density-dependent cluster model}",
  author={Ni, Dongdong and Ren, Zhongzhou},
  journal={Nucl. Phys. A},
  doi={10.1016/j.nuclphysa.2009.07.014},
  volume={828},
  number={3-4},
  pages={348--359},
  year={2009},
  publisher={Elsevier}
}

@article{yong2010,
  title="{Improvement of a fission-like model for nuclear $\alpha$ decay}",
  author={Yong-Jia, Wang and Hong-Fei, Zhang and Wei, Zuo and Jun-Qing, Li},
  journal={Chinese Phys. Lett.},
  doi={10.1088/0256-307X/27/6/062103},
  volume={27},
  number={6},
  pages={062103},
  year={2010},
  publisher={IOP Publishing}
}

@book{delion2010,
  title={Theory of particle and cluster emission},
  author={Delion, Doru S},
  volume={819},
  year={2010},
  publisher={Springer}
}

@article{lu2024,
  title = "{Effects of different effective nucleon-nucleon interactions on $\ensuremath{\alpha}$-decay half-life and extracted $\ensuremath{\alpha}$-cluster preformation probability}",
  author = {Lu, Mingzhao and Wan, Niu},
  journal = {Phys. Rev. C},
  volume = {110},
  issue = {4},
  pages = {044321},
  numpages = {9},
  year = {2024},
  month = {Oct},
  publisher = {American Physical Society},
  doi = {10.1103/PhysRevC.110.044321}
}

@article{shin2016,
  title = "{Nuclear isospin asymmetry in $\ensuremath{\alpha}$ decay of heavy nuclei}",
  author = {Shin, Eunkyoung and Lim, Yeunhwan and Hyun, Chang Ho and Oh, Yongseok},
  journal = {Phys. Rev. C},
  volume = {94},
  issue = {2},
  pages = {024320},
  numpages = {13},
  year = {2016},
  month = {Aug},
  publisher = {American Physical Society},
  doi = {10.1103/PhysRevC.94.024320}
}

@article{qian2011,
  title = "{Calculations of $\ensuremath{\alpha}$-decay half-lives for heavy and superheavy nuclei}",
  author = {Qian, Yibin and Ren, Zhongzhou and Ni, Dongdong},
  journal = {Phys. Rev. C},
  volume = {83},
  issue = {4},
  pages = {044317},
  numpages = {8},
  year = {2011},
  month = {Apr},
  publisher = {American Physical Society},
  doi = {10.1103/PhysRevC.83.044317}
}

@article{zhang2025,
  title = {Roles of spin-orbit interaction and pairing correlation in $\ensuremath{\alpha}$ decay of $^{210}\mathrm{Po}$},
  author = {Zhang, Wenhao and Gao, Yonghao and Cui, Jianpo and Wang, Yanzhao and Gu, Jianzhong},
  journal = {Phys. Rev. C},
  volume = {112},
  issue = {2},
  pages = {024319},
  numpages = {10},
  year = {2025},
  month = {Aug},
  publisher = {American Physical Society},
  doi = {10.1103/dw7k-fj9j}
}

@article{pfutzner2012,
  title = "{Radioactive decays at limits of nuclear stability}",
  author = {Pf\"utzner, M. and Karny, M. and Grigorenko, L. V. and Riisager, K.},
  journal = {Rev. Mod. Phys.},
  volume = {84},
  issue = {2},
  pages = {567--619},
  numpages = {0},
  year = {2012},
  month = {Apr},
  publisher = {American Physical Society},
  doi = {10.1103/RevModPhys.84.567}
}

@article{qi2019,
  title="{Recent developments in radioactive charged-particle emissions and related phenomena}",
  author={Qi, Chong and Liotta, Roberto and Wyss, Ramon},
  journal={Prog. Part. Nucl. Phys.},
  doi={10.1016/j.ppnp.2018.11.003},
  volume={105},
  pages={214--251},
  year={2019},
  publisher={Elsevier}
}

@article{denisov2024,
  title = "{Empirical relations for $\ensuremath{\alpha}$-decay half-lives: The effect of deformation of daughter nuclei}",
  author = {Denisov, V. Yu.},
  journal = {Phys. Rev. C},
  volume = {110},
  issue = {1},
  pages = {014604},
  numpages = {8},
  year = {2024},
  month = {Jul},
  publisher = {American Physical Society},
  doi = {10.1103/PhysRevC.110.014604}
}

@article{sun2016,
  title = "{Systematic study of $\ensuremath{\alpha}$ decay half-lives for even-even nuclei within a two-potential approach}",
  author = {Sun, Xiao-Dong and Guo, Ping and Li, Xiao-Hua},
  journal = {Phys. Rev. C},
  volume = {93},
  issue = {3},
  pages = {034316},
  numpages = {9},
  year = {2016},
  month = {Mar},
  publisher = {American Physical Society},
  doi = {10.1103/PhysRevC.93.034316}
}

@article{sun2017,
  title = "{Systematic study of $\ensuremath{\alpha}$ decay for odd-$A$ nuclei within a two-potential approach}",
  author = {Sun, Xiao-Dong and Duan, Chao and Deng, Jun-Gang and Guo, Ping and Li, Xiao-Hua},
  journal = {Phys. Rev. C},
  volume = {95},
  issue = {1},
  pages = {014319},
  numpages = {11},
  year = {2017},
  month = {Jan},
  publisher = {American Physical Society},
  doi = {10.1103/PhysRevC.95.014319}
}

@article{zheng2025,
  title = "{Effect of deformation on $\ensuremath{\alpha}$ clustering from a pocket-type dynamical double-folding potential}",
  author = {Zheng, Hailan and Deng, Daming and Ren, Zhongzhou},
  journal = {Phys. Rev. C},
  volume = {112},
  issue = {6},
  pages = {064320},
  numpages = {11},
  year = {2025},
  month = {Dec},
  publisher = {American Physical Society},
  doi = {10.1103/hqtq-xvyp}
}

@article{nhule2025,
title = "{Implications of repulsive core effects in double-folding potentials for $\alpha$ decay of heavy and superheavy nuclei}",
journal = {Nucl. Phys. A},
volume = {1063},
pages = {123217},
year = {2025},
issn = {0375-9474},
doi = {10.1016/j.nuclphysa.2025.123217},
author = {N. Nhu Le}
}

@article{deng2016,
  title = "{$\ensuremath{\alpha}$ preformation factors of medium-mass nuclei and the structural effects in the region of crossing the $Z=82$ shell}",
  author = {Deng, Daming and Ren, Zhongzhou},
  journal = {Phys. Rev. C},
  volume = {93},
  issue = {4},
  pages = {044326},
  numpages = {9},
  year = {2016},
  month = {Apr},
  publisher = {American Physical Society},
  doi = {10.1103/PhysRevC.93.044326}
}

\end{document}